\documentstyle[12pt,psfig,aaspp]{article}
\def\ge{G_e}

\def\grad{\nabla}

\def\beq{\begin{equation}}
\def\eeq{\end{equation}}
\begin{document}

\title{Dwarf satellite galaxies in the modified dynamics}
\author{ Rafael Brada and Mordehai Milgrom}
\affil{ Condensed Matter Department, Weizmann Institute of Science, Rehovot Israel}

\begin{abstract}

In the modified dynamics (MOND) the inner workings
of dwarf satellites can be greatly affected by their mother galaxy--over and 
beyond its tidal effects. Because of MOND's nonlinearity
 a system's internal
 dynamics can be altered by an external field
in which it is immersed (even when this field, by itself, is constant in
 space). As a result, the size and
velocity dispersion of the satellite  vary as the external field varies 
along its orbit. A notable outcome of this is a substantial increase in the 
dwarf's vulnerability to eventual tidal disruption--rather higher than 
 Newtonian dynamics (with a dark-matter halo) would lead us to expect for
a satellite with given observed parameters.
 The space of system parameters of the dwarf may be divided
according to three main criteria:
It can be either external- or internal-field dominated;
 it can be in the adiabatic
or in the impulsive regime; and 
it can be in the tidal or non-tidal regime.
 The Milky Way's dwarf satellites populate all these
regions of parameter space, and a single dwarf on an eccentric orbit can
traverse several of them. The situation is particularly transparent
 in the external-field-dominated, adiabatic
regime where the dynamics can be described analytically: due to the
variation in the external-field strength with the galactocentric distance of
 the dwarf, $R$,
its size changes as $R^{-1}$, and the internal velocities
change as $R$. As the dwarf approaches the center
it puffs up, becoming susceptible to tidal disruption.
  Adiabaticity is lost roughly at the same
$R$ were tidal effects become important.
The behavior near and in the impulsive regime is studied numerically.

\end{abstract}
%%%%%%%%%%%%%%%%%%%%%%%%%%%%%%%%%%%%%%%%%%%%%%%%%%%%%%%%%%%%%%%%%%%%%
\section{Introduction}
The dynamical behavior of dwarf spheroidals and other satellites  of the Milky Way holds much information
pertinent to the dark-matter problem. Attempts to elicit such  knowledge include, on the one
hand, measurements of the satellites' intrinsic properties such as the size, luminosity, and
velocity dispersion, which evince mass discrepancies in the satellites
 (\cite{ao,pryor91,matar}). This discrepancy is removed in the modified 
dynamics--MOND (\cite{dwarf,mdeb,matar}).
 On the other hand,
the satellites can be used to probe the gravitational field of their mother 
 galaxy (specifically, the Milky Way)
by using them as test particles to probe the galaxy's potential field
 (e.g., \cite{lt}), or by studying
 tidal effects of the galaxy on the structure of the 
satellite taken as
a finite body (e.g., \cite{fl}).
 In Newtonian dynamics, the history of the center-of-mass motion 
 may influence the internal workings of the satellite via tidal effects.
Tidal disruption may have culled from the satellite population those that
 are internally weakly
bound and/or move on elongated orbits, thus affecting the distribution of
 galactic orbits seen today (see, e.g., \cite{lcg}).
\par
 In MOND, the interaction between the internal and center-of-mass motions,
 brought about by the theory's nonlinearity,
 goes beyond the Newtonian effects. For small systems (smaller than
the scale over which the external field varies) the effect goes in one
 direction: while the center-of-mass motion is not affected by the internal
motions, it may strongly affect them as explained in \cite{bm} and 
\cite{mgfe}.
This occurs
when the accelerations inside the satellite are of the order of or smaller
 than its center-of-mass acceleration; it is also required that the internal 
accelerations be small compared with the acceleration constant of MOND,
 $a_0$, as is always the case for the Milky Way's dwarf satellites.
 Due to this external-field effect (EFE) a satellite that plunges into
 the galaxy on an eccentric  orbit increases
in size, making itself an easier victim for tidal disruption. 
An additional destructive
effect results when the changes in the external field become resonant
with the internal motions.
\par
The purpose of the paper is to describe, and demonstrate the pertinence,
of these processes, 
which  are peculiar to MOND.
\par
In the next section we briefly recapitulate the external field effect. Next,
in section 3, we consider a dwarf on an elongated orbit,
delineating
 the different regimes of application of the MOND effects, and give an
 analytic description of dynamics in the adiabatic regime. 
In section 4 we describe the MOND N-body simulations, the results of which are
described in section 5. Section 6 lists our conclusions and briefly comments
on the Milky Way's dwarf satellites.
%%%%%%%%%%%%%%%%%%%%%%%%%%%%%%%%%%%%%%%%%%%%%%%%%%%%%%%%%%%%%%%%%%%%%

\section{The External-Field Effect}
\label{sec:efe}
We work with the formulation of MOND as modified gravity  described in
\cite{bm}
 whereby the Poisson 
equation for the nonrelativistic gravitational potential is replaced by
 \beq \label{fe}\grad\cdot[\mu(|\grad\phi|/a_0)
\grad\phi]= 4\pi G\rho \label{field} \eeq 
(and the gravitational acceleration is given by $-\grad\phi$).
 For a small system, freely falling in 
an external field that dominates its own,
equation(\ref{fe}) can be linearized in the internal field, expanding about
the value of the external acceleration $g_{ex}$
 (approximately constant over the
 extent of the small system).
As shown in  \cite{mgfe} one gets a  quasi-Newtonian internal dynamics 
 with an effective gravitational constant $\ge=G/\mu(g_{ex}/a_0)$.
We use the term ``quasi-Newtonian'' because, in addition to the increased
 effective gravitational constant, the dynamics is anisotropic
 with some dilation along the direction of
the external field.
If the external field is in the $z$ direction then the linearized
 equation reduces to the Poisson equation in the coordinates
 $x,y,z(1+L)^{-1/2}$, where $L\equiv dln[\mu(s)]/dln(s)$ at $s=g_{ex}/a_0$
 takes a value between 0 and 1.
 In the 
deep-MOND limit [$s\ll 1$, where $\mu(s)\approx s$],
 assumed all along in this paper, we then have 
 \beq\ge=Ga_0/g_{ex}\gg G,   \label{zuma} \eeq
 and $L\approx 1$.
%%%%%%%%%%%%%%%%%%%%%%%%%%%%%%%%%%%%%%%%%%%%%%%%%%%%%%%%%%%%%%%%%%%
\section{A satellite on an elongated orbit}
\label{orbit}
MOND's basic premise is that our galaxy, like others, does not contain
dynamically important dark matter. Thus, as long as the orbit of the
 satellite under
consideration does not take it within a few scale lengths of the mother 
 galaxy, the latter may be treated
as a point mass. Even within galaxies the mean accelerations never much exceed
$a_0$; at large galactocentric distances 
 the acceleration is always smaller than
 $a_0$, as we assume all along.
 The MOND-limit acceleration at a distance $R$ from a point mass $M$
 is
\beq
g(R) = V_\infty ^2/R, \label{dosta}
\eeq
where $V_\infty\equiv (GMa_0)^{1/4} $ is the asymptotic,
 circular-orbit speed.
This $g(R)$ is the external acceleration field $g_{ex}$
 that enters the quasi-Newtonian internal
dynamics of the dwarf on the sections of its orbit where $g(R)$
outweighs the internal accelerations.
If along some portion of the dwarf's orbit 
the change in the external field
is slow--i.e., occurs on time scales long compared with the internal
 dynamical
time--the quantity $vr$ is expected to remain constant as an adiabatic 
invariant. Here $v$ is some mean internal velocity, and $r$ is the mean
radius of the system.
Also, in the 
quasi-Newtonian regime an effective, Newtonian virial relation should
hold: $ v^2 \approx \ge Mr^{-1}$. As $R$ varies along this section of the 
orbit, and with it $\ge$, we expect $v$ and $r$ to follow according to
 $ v \propto\ge\propto R $, and $ r \propto\ge^{-1}\propto R^{-1} $.
As the dwarf plunges in on an eccentric orbit it puffs up--an effect that
does not appear in Newtonian dynamics with dark matter--thus rendering itself
 more susceptible to tidal breakup than it would be due to the increasing
external-field gradients alone.
\par
To consider more quantitatively the interplay between adiabaticity,
 external-field dominance, and tidal breakup,
consider a satellite described by its gross properties:
the (baryonic)  mass $m$, the root-mean-square velocity of the constituents
 with respect to the center-of-mass $v$,
and the size $r$ (say the rms distance of constituents from the center). It
 moves on an orbit $R(t)$ with velocity $V(t)$ in the field of the 
point-like mother galaxy of mass $M$. (We neglect the secondary effects of 
anisotropy and so we only consider the magnitude of the position vector
 $\vec R$.)

The parameter
\beq\beta \equiv v^2/rg(R)= v^2 R /V^2_{\infty}r \label{beta}  \eeq
measures the importance of the internal acceleration vis-a-vis
 the external one. The EFE is pertinent  when $\beta \lesssim 1$.
In terms of the radii and masses we can write
\beq \beta\approx\cases{(R/r)(m/M)^{1/2}&if $\beta\gg 1$;\cr(R/r)^2(m/M)&if
$\beta\ll 1$.\cr} \label{klata} \eeq 
 Here we used Newtonian expressions
with $\ge$ from eq.(\ref{zuma}) when $\beta\ll 1$.
The parameter
 \beq \gamma \equiv (R/V_{\infty})/(r/v)
=(R/r)^{1/2}\beta^{1/2} \label {alpha}\eeq 
is useful for measuring the degree of adiabaticity
 (achieved when  $\gamma \gg 1$) when the orbit is mildly eccentric, because
 then the orbital changes occur on a time scale $R/V$,
 and $V\sim V_{\infty}$. 
 (The MOND potential far from a central mass is logarithmic, for which the
virial relation reads $\langle V^2\rangle=V^2_{\infty}$. The velocities
at perigalacticon, $V_p$, and apogalacticon, $V_a$, are related by
$V^2_p-V^2_a=V^2_{\infty}ln(R_a/R_p)$, where $R_p$ and $R_a$ are the 
respective distances.)
We can write
\beq \gamma\approx\cases{(R/r)(m/M)^{1/4}&if $\beta\gg 1$;\cr(R/r)^{3/2}
(m/M)^{1/2}&if  $\beta\ll 1$.\cr} \label{ktupa} \eeq 
 Because only  $R\gg r$ is of interest, we see from eq.(\ref{alpha}) that
$\gamma\gg 1$ in the whole region $\beta>1$.
\par
Tidal effects in the bulk are important when the mean internal acceleration, 
$g_{in}$, is
 smaller than the increment of the external acceleration over
the extent $r$; i.e., when $g_{in}\lesssim V_\infty^2 r/R^2=g(R)r/R$.
 We take as
 the criterion for the importance of tidal effects 
 \beq  \alpha\equiv [g_{in}/(V_\infty^2 r/R^2)]^{1/3}=(vR/V_{\infty}r)^{2/3}
\lesssim 1.
 \label{matut} \eeq
 Again,
since only cases for which $r\ll R$ are of interest we see that tidal 
effects need concern us only when $g_{in}\ll g$ ($\beta\ll 1$). In this regime
we have $\alpha\approx (R/r)(m/M)^{1/3}$. 
Note in general that
 $\alpha=\gamma^{2/3}$.
This means that non-adiabaticity and
tidal effects enter at about the same place on the orbit, as is
 indeed verified
 in our numerical calculations. Clearly, the inflation
of the satellite due to the EFE continues in the tidal phase. 
\par
We can now qualitatively see what happens to a dwarf on an elongated orbit.
If the entire orbit has $\beta\gg 1$ the satellite is 
unaffected by the Galaxy. If its orbit takes it to a small enough 
galactocentric distance $R_0$ where $\beta=1$ the EFE enters into action
there.
At this point the situation is adiabatic with
 $\gamma_0=\gamma(R_0)\approx(R_0/r_0)^{1/2}$, which is $\sim 10$ for the 
typical value of $R_0/r_0\sim 100$.
 As $R$ decreases further
we are, at first, in the adiabatic regime with $r\approx r_0 R_0/R$, 
 $v\approx v_0 R/R_0$, and $\beta$ decreasing still below 1: 
$\beta\approx (R/R_0)^4$.
In this region $\gamma\approx \gamma_0(R/R_0)^3$  from eq.(\ref{ktupa}), 
so roughly at 
$R= R_0\gamma_0^{-1/3}$ adiabaticity is lost, and at the same time 
tidal effects set in, in which case we have to resort to MOND, N-body
calculations, as described below.
\par
The comparison of the MOND predictions on the onset of tidal effects
 with those of Newtonian dynamics (ND) (with dark matter) depends
on what exactly is measured, and on what is assumed in ND (e.g., on the
 dark-matter distribution in the dwarf). But, in any event, the puffing up
of a dwarf in the $\beta\lesssim 1$ region, which has no analogue in ND,
makes dwarfs more vulnerable to tidal disruption. To take a specific 
example,  
suppose a satellite is observed at $R=R_1$ with measured size,
internal, and center-of-mass velocities. Its future orbit can 
then be deduced,
and also the Newtonian, dynamical mass it contains, $m_N$. Suppose
 it is already in the 
$\beta\lesssim 1$ regime. It is easy to see that the value
of its tidal parameter $\alpha$ as deduced in ND
 is the same as that given by MOND,
since $m_N=m\ge(R_1)/G$ and the galactic mass within $R_1$ is
$M_N(R_1)=M\ge(R_1)/G$. As we saw, MOND predicts that
 $\alpha\propto R^2$, while Newtonian dynamics predicts
$\alpha\propto R^{2/3}$, since $r$ is then assumed to remain constant
while $M_N(R)\propto R$. So tidal effects will clearly enter at larger
radii in MOND.

%%%%%%%%%%%%%%%%%%%%%%%%%%%%%%%%%%%%%%%%%%%%%%%%%%%%%%%%%%%%%%%%%%%%%%%%%
\section{N-body simulations}
The numerical simulations involve a model dwarf comprising N identical
particles that starts with some equilibrium distribution function
 in compliance with MOND dynamics.
The model is then subjected to different types of variable external influences
that mimic aspects of the influence of the mother galaxy. The underlying
potential field equation is eq.(\ref{field}), or its linearized, 
approximate form. This nonlinear potential equation is solved 
numerically using multi-grid methods as detailed
in  \cite{thesis} and adumbrated in \cite{stab}.
The particles are then propagated in the derived potential. It is only 
interesting to study the dwarf when it is in the external-field-dominated
region.
To isolate the different effects discussed above we proceed in three steps. 
First, to pinpoint the effects of non-adiabaticity, and verify our analytic
deductions for the adiabatic regime, we start with a quasi-Newtonian
King model for the dwarf, assume quasi-Newtonian dynamics, and simply
vary $\ge$ periodically and see how the model reacts for different
frequencies of the perturbation.
In the second step we still consider a quasi-Newtonian behavior but 
the applied variations in $\ge$ and the direction of the external field 
 correspond to actual orbits
of a dwarf. The third stage, which is more costly, is to simulate the
complete system of dwarf plus a point-mass galaxy.
Since the construction of initial models for the dwarf are peculiar to
MOND we describe them briefly now, referring the reader for more details
 to \cite{thesis}.

\subsection{Constructing steady-state galactic models}

 We use as initial states King models (For details see, 
e.g.,  \cite{bt}) properly modified to constitute, as the case may require,
quasi-Newtonian or deep-MOND steady states.
The distribution function for the Newtonian models is
\beq
\label{kingdf}
f_{K} =\cases{
  \rho_1 (2\pi\sigma^2)^{-3/2} (e^{\varepsilon/\sigma^2}-1) & if
  $ \varepsilon >0$ \cr
  0 & if $\varepsilon \leq 0$,}  
 \eeq
where $ \varepsilon \equiv -E + \phi_{0} $, $E=v^2/2+\phi$, and
the parameter $\phi_{0}$ is the upper energy cutoff.
Equation(\ref{kingdf}) is integrated over velocities to obtain the 
density $\rho_{K}(\Psi)$
 as a function of the relative potential $\Psi \equiv -\phi+\phi_{0}$.
Instead of the Poisson equation we solve here the spherically symmetric
MOND equation (in the deep-MOND limit assumed all along):
\beq \label{poiss}
[a_0^{-1}r^2 (\Psi')^2]'=-4\pi G\rho_{K}(\Psi) \eeq
(the apostrophe signifies derivative with respect to $r$),
which provides an ordinary differential equation for $\Psi(r)$
that can be integrated numerically with the boundary
condition
 $\Psi'(0)=0$. The second boundary condition is
$\Psi(0)$, which together with $\phi_0$ determines the model. 
The model can also be 
specified in terms of other parameters from among the  
tidal radius $r_t$, the total mass, the central density $\rho(0)$,
and the King radius 
 $r_0 \equiv (9 \sigma^2/4 \pi G \rho_0)^{1/2}$. Note that for MOND King
 models $r_0$ defined in this way is not some characteristic radius; it is just a convenient representation of $\rho_0$ (so we can have $r_0>r_t$,
 for example).

%\beq
%\label{eq:monforp}
%\frac{d}{dr} \left[ \frac{r^2}{a_0} \left( \frac{d\Psi}{dr}\right)^2\right]=
%4\pi G\rho_1r^2\left[
%\: e^{\Psi/\sigma^2} \: erf\left(\frac{\Psi^{1/2}}{\sigma}\right)
%-\left(\frac{4 \Psi}{\pi \sigma^2}\right)^{1/2}\left(1+\frac{2 \Psi}{3
% \sigma^2} \right) \right]
%\eeq
In constructing a quasi-Newtonian model we remember that the transformed
 potential $\phi'(x',y',z')\equiv\phi(x,y,z)$, with 
 $x'=x,~y'=y,~z'=z(1+L)^{-1/2}$, 
with $L=dln[\mu(s)]/dln(s)$,
satisfies the usual Poisson equation with $\ge$ as gravitational constant, and 
 $\rho'(x',y',z')\equiv\rho(x,y,z)$ as density ($z$ is taken in the direction
of the external field). We thus begin by constructing a Newtonian model in the
auxiliary coordinates
$x',~x',~z'$ remembering that the conserved quantity on which the distribution
function depends by the Jeans theorem 
is $[v_x^{'2} + v_y^{'2} + (1+L) v_z^{'2}]/2  +   \phi'(x',y',z')$. This
 model has a spherical mass distribution, but a 
$v'_z$ dispersion that is smaller by a factor $(1+L)^{1/2}$ than those
in  the other
 directions. We draw positions and velocities for the auxiliary coordinates of
the N particles. Then we multiply all $z'$ and $v'_z$ values by  $(1+L)^{1/2}$.
The total mass of the model is multiplied by the same factor because
$\int\rho'(\vec r')d^3r'=(1+L)^{-1/2}\int\rho(\vec r)d^3r$.
The resulting model is elongated in the $z$ direction and has an
isotropic global velocity distribution.

%%%%%%%%%%%%%%%%%%%%%%%%%%%%%%%%%%%%%%%%%%%%%%%%%%%%%%%%%%%%%%%%%%

\section{Results of the simulations}

\subsection{Testing for the external-field effect}
\begin{figure}[htp]
\centerline{ \psfig{figure=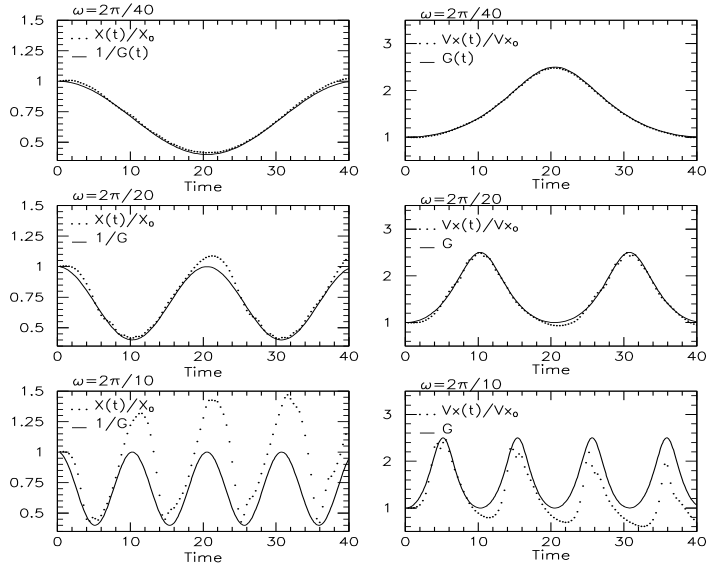,height=8cm,width=10cm}}
%\centerline{ \psfig{figure=fig1.eps,width=10cm}}
\caption{\protect\small The time dependence of $X(t)/X(0)$ and
$\ge(t)^{-1}$ (left-hand panels), and  $V_x(t)/V_x(0)$ and $\ge(t)$
 (right-hand
panels) for a quasi-Newtonian King model with  $m=1.02$, $X(0)=0.44$,
$V_x(0)=0.39$ and $\ge=(0.7+0.3 \cos\omega t)^{-1}$.
The quantities plotted in each panel should be equal for strict adiabaticity. 
The values of  $\omega$ are marked.}
\label{fig1}
\end{figure}

We first want to establish the consequences of the external-field effect and learn
what is the time scale necessary for a change to be adiabatic.
As was discussed in section \ref{orbit}, in the adiabatic regime we expect
 the average velocity in a system,
$v$, to be proportional to $\ge=G(a_0/g_{ex})$, and the average size
 of the system to
be inversely proportional to $\ge$.
 We start with a Newtonian
 King model having $10^5$ particles with
$ \sigma^2=1$, $r_0=1$, and $ \Psi(0)/\sigma^2=1$. We also take $G=1$ [this
fixes $\rho(0)$; the  total mass of the Newtonian
model is then $m=0.72$, and its tidal radius $r_t=1.975$]. We then perform the
stretching by $2^{1/2}$ (we take $L=1$ for the deep-MOND limit) to get 
a model with
$m=1.02 $ and root-mean-square
 values of the coordinate and velocity components
(designated by capital letters) 
 $X=Y=0.44 $, $Z=0.62 $,
$ V_{x}=V_{y}=V_{z}=0.39 $.
The natural dynamical time $r/v$ is of order unity.
We ran simulations on a cubical grids with $129^3$ grid
points using the quasi-Newtonian field equation with a fixed time step $dt=0.04$ for $10^3$
time steps. We varied $\ge$ periodically with time taking
$\ge(t)=(0.7+0.3 \cos\omega  t)^{-1}$ for
 three values of 
 $\omega  = 2 \pi(1/40,~1/20,~1/10) $.
( The physical
grid spacing is changed in proportion to $\ge^{-1}$ during the simulation,
 since we expect $r$ to scale as $\ge^{-1}$.)
In analogy with the adiabaticity parameter, $\gamma$, defined 
locally on an 
orbit, we can define here as some global measure of adiabaticity
$\hat\gamma\equiv (T/4)/[X(0)/V_x(0)]$, where $T=2\pi/\omega$ is the period.
 For the three models presented 
 $\hat\gamma= 10,~5,~2.5$ respectively.
 The results of these 
simulations for the $x$ components are shown in figure (\ref{fig1}) describing
how the extent and velocity dispersion in the $x$ direction vary with time. 
The results for the $z$ direction are the same within a few percents. 
The results for $\omega= 2 \pi/40$  show the expected
external-field effect with strict adiabaticity;
small departures from adiabaticity appear in the time dependence of the 
size and mean velocity for $\omega = 2 \pi/20$;
while for $\omega = 2 \pi /10 $ clear departure from adiabaticity is evident.
\par
We also learn that departure from adiabaticity brings about
a secular increase in the radius and decrease in the 
internal velocities.

%------------------------------------------------------

\subsection{The evolution of a quasi-Newtonian model along a realistic orbit}
Before going to the more complete models that utilize the full MOND field
 equation
and include the tidal forces, we follow the evolution of a quasi-Newtonian
 model varying the value of $\ge$ and the
 direction of the external field according to
 the location of the dwarf on an actual orbit in the logarithmic potential 
of the point-mass mother galaxy. Tidal forces
are then not taken into account.
 The purpose of these experiments is to test the degree to which the
changes in dwarf characteristics are adiabatic
 for sample models with realistic parameters. In particular,
 to see  what lasting effects 
non-adiabaticity near perigalacticon has on the dwarf in disjunction from 
tidal effects.
 We thus took orbits with strong adiabaticity
at apogalacticon, where we start, but a breakdown of adiabaticity near
 perigalacticon. Note that $\ge$ goes back to the same value at subsequent 
apogalacticons; so, apart from
some remaining oscillations on the dynamical time scale,
 the virial relation is reestablished in the mean, and, thus,
 $v^2r$ must come back to the same value.
\par
 We take the mass of the mother
galaxy as unit, $M=1$. Since we also use units in which $G=1$ and $a_0=1$,
the unit of velocities becomes
 $V_\infty\equiv (MGa_0)^{1/4}=1$, which in cgs units 
is about $220 kms^{-1}$ for the MW.
Length is then measured in units of $V_\infty^2/a_0$,
about 10 kpc for the MW. In these units the MW satellites are  typically
 at distances between  5 and 20, of size $3\times 10^{-2}$ to $10^{-1}$,
velocity dispersion $2\times 10^{-2}$ to $5\times 10^{-2}$,
and of mass $10^{-5}$ to $10^{-4}$.
Accordingly, we construct our dwarf model as a
 quasi-Newtonian King model having the following properties:
$\sigma^2=8\times 10^{-4}$, $ r_0=1/16 $, $\Psi(0)/\sigma^2=1$, $r_t=0.1234$
 (before  stretching), and
$m=5.09\times 10^{-5}$. We simulated its evolution for two orbits with 
pericenter and apocenter distances of
$R_{min}=6$, $R_{max}=9$, and $R_{min}=6$, $R_{max}=12$.
At galactocentric distance $R$ the external acceleration
 is $R^{-1}$ (in our units of $a_0$), and $\ge=R$.
\begin{figure}[htp]
\centerline{ \psfig{figure=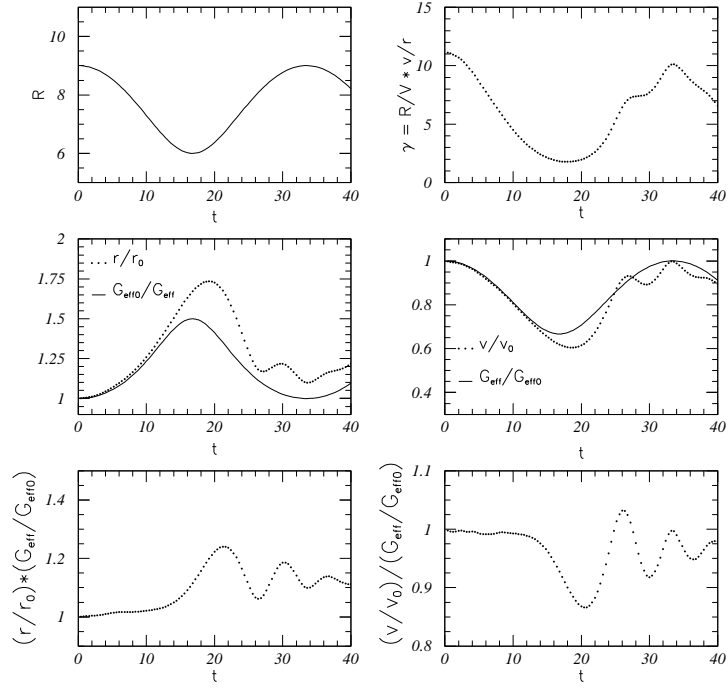,width=10cm}}
\caption{\protect\small  Orbital and intrinsic parameters as functions of 
time for a dwarf subject to an EFE. The field varies as it would on
 an orbit with $R_{max}=9 $ $R_{min}=6$ (tidal effects are excluded).}
\label{fig2}
\end{figure} 

\begin{figure}[htp]
\centerline{ \psfig{figure=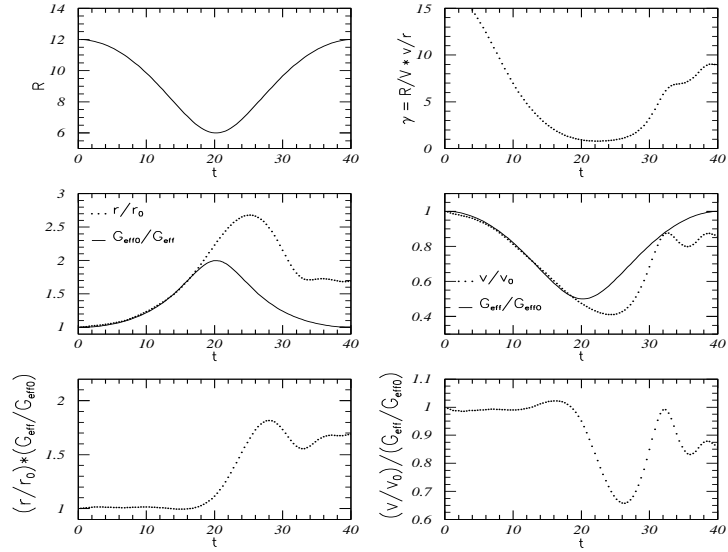,height=8cm,width=10cm}}
\caption{\protect\small Same as figure 2 for $R_{max}=12$ and $R_{min}=6$.}
\label{fig3}
\end{figure}

The results of the simulations are summarized in figures
 (\ref{fig2})(\ref{fig3}).
  Adiabaticity is more severely violated 
in the  second model, which has $R_{max}=12$. We see that, as a result
 of violating
adiabaticity near $R=R_{min}$,  the dwarf still oscillates on the dynamical
 time scale when it next enters the adiabatic regime around apocenter.
More importantly, it attains a larger radius (averaged over the fluctuations).
 The velocity
dispersion is correspondingly smaller ($v^2r$ is preserved and $vr$ is larger).
On the next close passage the dwarf will be even less adiabatic and
becomes more vulnerable and will continue to increase in size.

\subsection{The evolution of a full MOND model along a realistic orbit.}
We then followed the full evolution of a dwarf obeying MOND orbiting  a
point mother
galaxy. We started by producing an isolated MOND King model with the following 
parameters: $\sigma^2=8.53\times 10^{-3}$, $r_0=0.3$ (remember that $r_0$ is
 just a proxy for $\rho_0$, not a characteristic radius), 
and $\Psi(0)/\sigma^2=1$. The resulting
model has a total mass $m=3.59\times 10^{-5}$ and a tidal radius $r_t=0.1346$.
Note that $\sigma$ is not the mean velocity dispersion of the MOND model.
 This can be gotten from the deep-MOND virial relation (\cite{lsy}),
 which in our units
 reads $\langle v^2\rangle =2m^{1/2}/3$, where $\langle v^2\rangle$ is the 
three-dimensional rms velocity. So we get for the one-dimensional rms velocity
$\langle v^2_x\rangle=1.33\times 10^{-3}$.

 These
global parameters are similar to the ones of the quasi-Newtonian model
 we have used in the
previous subsection, but the density profiles of the two models differ: the
MOND model is less concentrated than the quasi-Newtonian model.
Our model dwarf is put on an eccentric orbit starting at an
 apogalacticon distance of
 $R_{max}=12$ and reaching a perigalacticon distance of $R_{min}=6$.
 From the above  model 
we construct a family of five models by scaling up the 
mass of the model ending up with masses $m$, $2m$, $3m$, $4m$, and $16m$,
and scaling the velocity dispersions up accordingly. These are all models for 
isolated MOND dwarfs.
However, at apocenter already we have to start with 
models under some external-field influence. So, before we let the
models evolve along the orbit we need to switch on adiabatically the external
 field. This is done in a preliminary simulation where we 
gradually increase the mass of the mother galaxy from zero to one.
\par
\begin{figure}[htp]
\centerline{ \psfig{figure=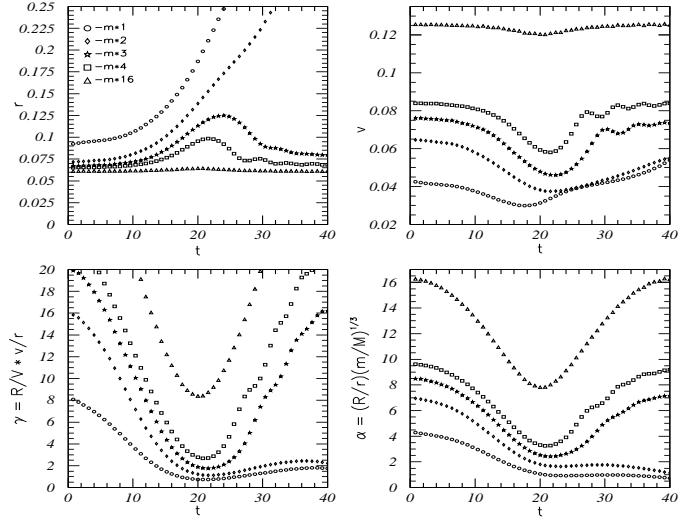,height=8cm,width=10cm}}
\caption{\protect\small Dwarf parameters as functions of time
 for five dwarf masses for the full model calculation,
all starting at apocenter at $R=12$ and reaching pericenter at $R=6$.}
\label{fig4}
\end{figure}
\begin{figure}[htp]
\centerline{ \psfig{figure=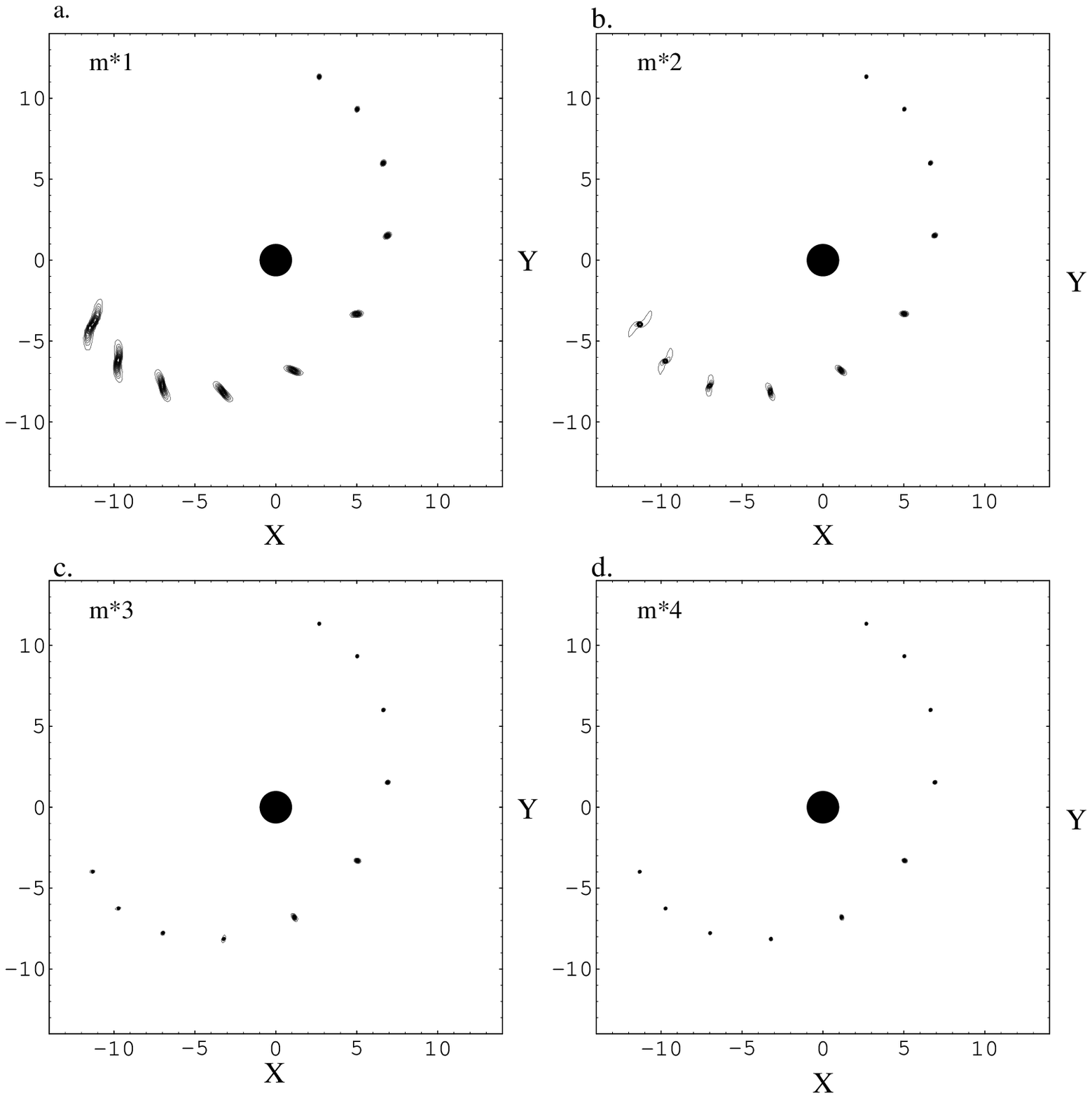,width=10cm}}
\caption{\protect\small  Projection of the mass on the $X-Y$ plane
 for the $1m$ (a), $2m$ (b), $3m$ (c), and  $4m$ (d) models. 
The mother galaxy is represented by the dark disk. Snapshots are taken every 4
time units (about every $2\times 10^8$ years for the MW).}
\label{fig5}
\end{figure}

The presence of the mother galaxy enters these
 simulations through the boundary conditions
used by the potential solver. The $16m$ model hardly changes when the external field is switched
on, while  for the  $m$ model the rms radius, $r$,
 increases by as much as $50\%$  when the external
field is switched on. The values for the parameter $\beta$, which  measures
 the ratio
of the internal field to the external field, calculated at $R=12$ are
 $0.233$, $0.7$, $1.03$, $1.29$, and $3.09$, respectively.
Thus, the $m$ model is dominated by the external field while
the $3m$ is the borderline case.
We then integrate the internal dynamics and the center-of-mass motion of each
of the models on the orbit. The orbits lie in the
$X$,$Y$ plane and we start at
$Y=12$, $X=0$. The simulations lasted for 40 time units and consisted of 
$10^3$ individual
time steps. The 
 rms value of $r$ and $v$ as functions of time are given in
 figure (\ref{fig4}). Also shown there are the adiabaticity ($\gamma$) 
and the tidal ($\alpha$) parameters along the
orbit. Projections of the dwarf structure  for the four smaller-mass models,
on the $X-Y$ plane, are given in 
figure (\ref{fig5}).

The value of the parameter $\beta$ at pericenter for the five models was:
$0.0315$, $0.056$, $0.168$, $0.2106$, and $1.36$, respectively.
The three models with masses $m$, $2m$ and $3m$ show clear signs of tidal disruption.
The $m$ model seems to have been totally destroyed by the tidal forces and there is no
clear core that remained after the passage near the galaxy. The $2m$ model was strongly
influenced by the tidal forces and lost about $25\%$ of its mass. The $3m$ model lost
only a few percent of its mass through tidal interaction.
We can attribute these mass losses to the combined action of tidal forces
and the extra non-adiabatic expansion of the models near $R=R_{min}$.
\par
In comparison, Newtonian dynamics applied to dwarfs observed with the same 
initial positions, center-of-mass velocities, sizes, and velocity dispersions
would predict much less tidal disruption. Consider, in particular, the two
most vulnerable models with masses $m$ and $2m$.  They start at $R=12$ 
with $\beta<1$, so, from the discussion at the end of section 3 we see that
their Newtonian $\alpha$ values there 
are the same as the MOND values. (Of course,
a Newtonist will assume that they contain more mass for the same
sizes and velocity dispersions.) From figure 4 we see that the two models
 start with $\alpha\approx 4,7$ respectively. The MOND scaling 
($\alpha\propto R^2$ for
 $\beta\ll 1$) implies that the two models should have $\alpha\approx 1, 1.7$
at perigalacticon, $R=6$, as they approximately do. In Newtonian dynamics,
 where $\alpha\propto R^{2/3}$, we would get at perigalacticon
$\alpha\approx 2.5, 4.4$ for the two models, making these initial 
model dwarfs much safer from later tidal disruption.
It need perhaps be clarified that the Newtonist will continue to get the same
 $\alpha$ values as in MOND if he uses at every point the observed
 properties, but this will lead him to conclude that the mass of the dwarf
 varies. Here we speak of what the Newtonist's predictions will be given
only the initial data, and assuming that the dwarf mass is constant.
\section{Summary and conclusions}
We have studied the existence, the nature, and the influence on dwarf
 satellites of the external field effect in MOND.
 For dwarf parameters in the EFE regime two situation are grossly
distinguished: a) the adiabatic regime, in which
tidal effects are not so important and b) the impulsive region, which also 
roughly coincides with the region where
 tidal forces become important. Due to the EFE the radius of a dwarf in 
the adiabatic regime increases as it approaches the mother galaxy.
If the whole orbit is in the adiabatic regime, the structure of the 
dwarf simply changes periodically with the orbital period.
If, however, some segment of the orbit is in the
impulsive-tidal regime near
 pericenter, then the dwarf might lose much of its mass there.
Even if it does not, it can emerge from this region
having a larger radius and smaller velocity dispersion (hence,
a longer intrinsic dynamical time). In its next approach to perigalacticon
it will thus enter the impulsive-tidal regime at a larger distance
from the center. 
\par
Clearly, all the above is highly germane to the dwarf system of the MW.
The distribution of intrinsic and orbital parameters of presently observed
 dwarfs must have been greatly affected by interaction with the 
MW. And, one
expects, MOND would give a different answer than  Newtonian 
dynamics with dark matter. To actually deduce the present-day properties
of the dwarfs would, however, require knowledge
of the initial distribution of the orbital and intrinsic parameters
of the dwarf-satellite population. Nothing is really known about this, so we
refrain from speculating on the subject.
 We only estimate where 
 our dwarf satellites stand
 as regards external-field dominance, adiabaticity,
and the importance of tidal effects.
\par
We consider the 10 dwarf spheroidal satellites
with known parameters (\cite{matar}): Sculptor, LSG 3, Fornax, Carina, 
Leo I, Sextans, Leo II, Ursa Minor, Draco, and Sagittarius.
 We take for the MW $V_{\infty}= 220$ kms$^{-1}$.
 Since only core radii, $r_c$, are given we write for the mean radius
$r=\eta r_c$ to get for
 the adiabaticity parameter of those dwarfs
$\gamma \sim \eta^{-1}(22,150,14,15,47,8,39,14,20,2)$, respectively. So, with the
exception of Sagittarius--which is known to be in the throes of
 disruption--and perhaps Sextans, 
these dwarfs are in the adiabatic regime within reasonable margins for $\eta$,
and even considering the approximate nature of the $\gamma$ criterion.
 According to our 
analysis they are also only weakly affected by tidal forces at their
 present positions.
 As has been pointed out
 (\cite{dwarf}, \cite{mdeb}), most of the above dwarfs
(with the exception of LSG 3, Leo I, and Leo II) are materially 
affected by the EFE: with the above choices of system parameters we 
get $\beta\sim \eta^{-1}(0.7, 4.4, 0.7, 0.5, 1.9, 0.2, 1.2, 0.6, 0.9, 0.1)$.
\par
If we apply the MOND scaling $\alpha\propto R^2$, which is valid
in the $\beta\ll 1$ regime, to the dwarfs with $\eta\beta < 1$
(except for Sagittarius) we can estimate
the minimum galactocentric distance
 above which the bulk of the dwarf is immune to tidal effects.
This is given by $R_t^M\approx R_0 \alpha_0^{-1/2}=R_0 \gamma_0^{-1/3}$,
 where here a subscript 0
marks present-day values. (If a dwarf in now on an outgoing section of its
orbit it will return to the same $R$, as it goes in, in the same state.)
For Sculptor, Fornax, Carina, Sextans, UMi, and Draco we get, respectively
$R_t^M\sim \eta^{1/3}(28, 57, 41, 43, 27, 32)$ kpc.
  The corresponding
 Newtonian values ($R_t^N\approx R_0\gamma_0^{-1}$) are
$R_t^N\sim \eta(4, 10, 7, 11, 5, 4)$ kpc. They are smaller than the
corresponding MOND values if $\eta$ is not so large that $\alpha_0<1$.
(For some dwarfs these Newtonian radii may fall within
the stellar MW where our approximation of a spherical, logarithmic potential
 is not valid.)
\par 
Our results imply 
that for a given dwarf in the adiabatic regime on an elongated orbit
under a strong EFE
the size and velocity dispersion would be strongly dependent on the distance
from the mother galaxy. One might then try to look for 
such correlations in the time-frozen population as seen today.
This seems to us quite hopeless at present because the effects will be swamped
by other factors of which we know very little;
 in particular, the unknown distribution of initial (intrinsic and orbital)
 parameters for the 
dwarfs. This is aggravated by the small sample size.  
\par
We leave for a future publication some other interesting effects predicted by
MOND that result from the EFE.
For example, in a dwarf in the EFE regime the total angular momentum is not 
conserved. We alluded to the fact that the direction of the external field
if felt by the ``internal'' dynamics of the dwarf. In a static or adiabatic
situation only the angular momentum along the external-field direction is
conserved.
\par
We thank the referee, Tad Pryor, for many useful comments and suggestions

\end{document}